\documentclass[twocolumn]{aastex631}

\newcommand{\src}{SGR 1806--20}
\newcommand{\srcshort}{SGR1806}
\newcommand{\srcone}{4U 0142+61}
\newcommand{\srctwo}{1RXS~J170849.0--400910}
\newcommand{\srctwosh}{1RXS~J1708}
\newcommand{\ixpe}{{\rm IXPE}}
\newcommand{\ixpeobssim}{\textsc{ixpeobssim}}
\newcommand{\xmm}{{\rm XMM-Newton}}
\newcommand{\nustar}{{\rm NuSTAR}}
\newcommand{\mdp}{\mathrm{MDP}_{99}}

\begin{document}
\AuthorCollaborationLimit=400

\title{\ixpe\ and \xmm\ observations of the Soft Gamma Repeater \src }

\correspondingauthor{Roberto Turolla}
\email{turolla@pd.infn.it}

\author[0000-0003-3977-8760]{Roberto Turolla}
\affiliation{Dipartimento di Fisica e Astronomia, Universit\`{a} degli Studi di Padova, Via Marzolo 8, I-35131 Padova, Italy}
\affiliation{Mullard Space Science Laboratory, University College London, Holmbury St Mary, Dorking, Surrey RH5 6NT, UK}

\author[0000-0002-1768-618X]{Roberto Taverna}
\affiliation{Dipartimento di Fisica e Astronomia, Universit\`{a} degli Studi di Padova, Via Marzolo 8, I-35131 Padova, Italy}

\author[0000-0001-5480-6438]{Gian Luca Israel}
\affiliation{INAF Osservatorio Astronomico di Roma, via Frascati 33, I-00078 Monteporzio Catone, Italy}

\author[0000-0003-3331-3794]{Fabio Muleri}
\affiliation{INAF Istituto di Astrofisica e Planetologia Spaziali, Via del Fosso del Cavaliere 100, I-00133 Roma, Italy}

\author[0000-0001-5326-880X]{Silvia Zane}
\affiliation{Mullard Space Science Laboratory, University College London, Holmbury St Mary, Dorking, Surrey RH5 6NT, UK}

\author[0000-0002-4576-9337]{Matteo Bachetti}
\affiliation{INAF Osservatorio Astronomico di Cagliari, Via della Scienza 5, I-09047 Selargius (CA), Italy}

\author[0000-0001-9739-367X]{Jeremy Heyl}
\affiliation{Department of Physics and Astronomy, University of British Columbia, Vancouver, BC V6T 1Z1, Canada} 

\author[0000-0003-0331-3259]{Alessandro Di Marco}
\affiliation{INAF Istituto di Astrofisica e Planetologia Spaziali, Via del Fosso del Cavaliere 100, I-00133 Roma, Italy}

\author[0000-0002-5250-2710]
{Ephraim Gau}
\affiliation{Physics Department and McDonnell Center for the Space Sciences, Washington University in St. Louis, St. Louis, MO 63130, USA}

\author[0000-0002-1084-6507]{Henric Krawczynski}
\affiliation{Physics Department and McDonnell Center for the Space Sciences, Washington University in St. Louis, St. Louis, MO 63130, USA}

\author[0000-0002-0940-6563]{Mason Ng}
\affiliation{MIT Kavli Institute for Astrophysics and Space Research, Massachusetts Institute of Technology, 77 Massachusetts Avenue, Cambridge, MA 02139, USA}

\author[0000-0001-5902-3731]{Andrea Possenti}
\affiliation{INAF Osservatorio Astronomico di Cagliari, Via della Scienza 5, I-09047 Selargius (CA), Italy}

\author[0000-0002-0983-0049]{Juri Poutanen}
\affiliation{Department of Physics and Astronomy, FI-20014 University of Turku, Finland}

\author[0000-0002-9785-7726]{Luca Baldini}
\affiliation{Istituto Nazionale di Fisica Nucleare, Sezione di Pisa, Largo B. Pontecorvo 3, I-56127 Pisa, Italy}
\affiliation{Dipartimento di Fisica, Universit\`{a} di Pisa, Largo B. Pontecorvo 3, I-56127 Pisa, Italy}

\author[0000-0002-2152-0916]{Giorgio Matt}
\affiliation{Dipartimento di Matematica e Fisica, Universit\`{a} degli Studi Roma Tre, Via della Vasca Navale 84, I-00146 Roma, Italy}

\author[0000-0002-6548-5622]{Michela Negro}
\affiliation{University of Maryland, Baltimore County, Baltimore, MD 21250, USA}
\affiliation{NASA Goddard Space Flight Center, Greenbelt, MD 20771, USA}
\affiliation{Center for Research and Exploration in Space Science and Technology, NASA/GSFC, Greenbelt, MD 20771, USA}





\author[0000-0002-3777-6182]{Iv\'{a}n Agudo}
\affiliation{Instituto de Astrof\'{i}sica de Andaluc\'{i}a-CSIC, Glorieta de la Astronom\'{i}a s/n, E-18008, Granada, Spain}

\author[0000-0002-5037-9034]{Lucio A. Antonelli}
\affiliation{Space Science Data Center, Agenzia Spaziale Italiana, Via del Politecnico snc, I-00133 Roma, Italy}
\affiliation{INAF Osservatorio Astronomico di Roma, Via Frascati 33, I-00078 Monte Porzio Catone (RM), Italy}

\author[0000-0002-5106-0463]{Wayne H. Baumgartner}
\affiliation{NASA Marshall Space Flight Center, Huntsville, AL 35812, USA}

\author[0000-0002-2469-7063]{Ronaldo Bellazzini}
\affiliation{Istituto Nazionale di Fisica Nucleare, Sezione di Pisa, Largo B. Pontecorvo 3, I-56127 Pisa, Italy}

\author[0000-0002-4622-4240]{Stefano Bianchi}
\affiliation{Dipartimento di Matematica e Fisica, Universit\`{a}  degli Studi Roma Tre, Via della Vasca Navale 84, I-00146 Roma, Italy}

\author[0000-0002-0901-2097]{Stephen D. Bongiorno}
\affiliation{NASA Marshall Space Flight Center, Huntsville, AL 35812, USA}

\author[0000-0002-4264-1215]{Raffaella Bonino}
\affiliation{Dipartimento di Fisica, Universit\`{a}  degli Studi di Torino, Via Pietro Giuria 1, I-10125 Torino, Italy}
\affiliation{Istituto Nazionale di Fisica Nucleare, Sezione di Torino, Via Pietro Giuria 1, I-10125 Torino, Italy}

\author[0000-0002-9460-1821]{Alessandro Brez}
\affiliation{Istituto Nazionale di Fisica Nucleare, Sezione di Pisa, Largo B. Pontecorvo 3, I-56127 Pisa, Italy}

\author[0000-0002-8848-1392]{Niccol\`{o} Bucciantini}
\affiliation{INAF Osservatorio Astrofisico di Arcetri, Largo Enrico Fermi 5, I-50125 Firenze, Italy}
\affiliation{Dipartimento di Fisica e Astronomia, Universit\`{a}  degli Studi di Firenze, Via Sansone 1, I-50019 Sesto Fiorentino (FI), Italy}
\affiliation{Istituto Nazionale di Fisica Nucleare, Sezione di Firenze, Via Sansone 1, I-50019 Sesto Fiorentino (FI), Italy}

\author[0000-0002-6384-3027]{Fiamma Capitanio}
\affiliation{INAF Istituto di Astrofisica e Planetologia Spaziali, Via del Fosso del Cavaliere 100, I-00133 Roma, Italy}

\author[0000-0003-1111-4292]{Simone Castellano}
\affiliation{Istituto Nazionale di Fisica Nucleare, Sezione di Pisa, Largo B. Pontecorvo 3, I-56127 Pisa, Italy}

\author[0000-0001-7150-9638]{Elisabetta Cavazzuti}
\affiliation{Agenzia Spaziale Italiana, Via del Politecnico snc, I-00133 Roma, Italy}

\author[0000-0002-4945-5079]{Chieng-Ting Chen}
\affiliation{Universities Space Research Association (USRA)} 
\affiliation{NASA Marshall Space Flight Center, Huntsville, AL 35812, USA} 

\author[0000-0002-0712-2479]{Stefano Ciprini}
\affiliation{Space Science Data Center, Agenzia Spaziale Italiana, Via del Politecnico snc, I-00133 Roma, Italy}
\affiliation{Istituto Nazionale di Fisica Nucleare, Sezione di Roma Tor Vergata, Via della Ricerca Scientifica 1, I-00133 Roma, Italy}

\author[0000-0003-4925-8523]{Enrico Costa}
\affiliation{INAF Istituto di Astrofisica e Planetologia Spaziali, Via del Fosso del Cavaliere 100, I-00133 Roma, Italy}

\author[0000-0001-5668-6863]{Alessandra De Rosa}
\affiliation{INAF Istituto di Astrofisica e Planetologia Spaziali, Via del Fosso del Cavaliere 100, I-00133 Roma, Italy}

\author[0000-0002-3013-6334]{Ettore Del Monte}
\affiliation{INAF Istituto di Astrofisica e Planetologia Spaziali, Via del Fosso del Cavaliere 100, I-00133 Roma, Italy}

\author[0000-0002-5614-5028]{Laura Di Gesu}
\affiliation{Agenzia Spaziale Italiana, Via del Politecnico snc, I-00133 Roma, Italy}

\author[0000-0002-7574-1298]{Niccol\`{o} Di Lalla}
\affiliation{Department of Physics and Kavli Institute for Particle Astrophysics and Cosmology, Stanford University, Stanford, CA 94305, USA}

\author[0000-0002-4700-4549]{Immacolata Donnarumma}
\affiliation{Agenzia Spaziale Italiana, Via del Politecnico snc, I-00133 Roma, Italy}

\author[0000-0001-8162-1105]{Victor Doroshenko}
\affiliation{Institut f\"{u}r Astronomie und Astrophysik, Universit\"{a}t T\"{u}bingen, Sand 1, D-72076 T\"{u}bingen, Germany}

\author[0000-0003-0079-1239]{Michal Dov\v{c}iak}
\affiliation{Astronomical Institute of the Czech Academy of Sciences, Bo\v{c}n\'{i} II 1401/1, 14100 Praha 4, Czech Republic}

\author[0000-0003-4420-2838]{Steven R. Ehlert}
\affiliation{NASA Marshall Space Flight Center, Huntsville, AL 35812, USA}

\author[0000-0003-1244-3100]{Teruaki Enoto}
\affiliation{RIKEN Cluster for Pioneering Research, 2-1 Hirosawa, Wako, Saitama 351-0198, Japan}

\author[0000-0001-6096-6710]{Yuri Evangelista}
\affiliation{INAF Istituto di Astrofisica e Planetologia Spaziali, Via del Fosso del Cavaliere 100, I-00133 Roma, Italy}

\author[0000-0003-1533-0283]{Sergio Fabiani}
\affiliation{INAF Istituto di Astrofisica e Planetologia Spaziali, Via del Fosso del Cavaliere 100, I-00133 Roma, Italy}

\author[0000-0003-1074-8605]{Riccardo Ferrazzoli}
\affiliation{INAF Istituto di Astrofisica e Planetologia Spaziali, Via del Fosso del Cavaliere 100, I-00133 Roma, Italy}

\author[0000-0003-3828-2448]{Javier A. Garcia}
\affiliation{California Institute of Technology, Pasadena, CA 91125, USA}

\author[0000-0002-5881-2445]{Shuichi Gunji}
\affiliation{Yamagata University, 1-4-12 Kojirakawa-machi, Yamagata-shi 990-8560, Japan}

\author{Kiyoshi Hayashida}
\altaffiliation{Deceased}
\affiliation{Osaka University, 1-1 Yamadaoka, Suita, Osaka 565-0871, Japan}

\author[0000-0002-0207-9010]{Wataru Iwakiri}
\affiliation{
International Center for Hadron Astrophysics, Chiba University, Chiba 263-8522, Japan}

\author[0000-0001-6158-1708]{Svetlana G. Jorstad}
\affiliation{Institute for Astrophysical Research, Boston University, 725 Commonwealth Avenue, Boston, MA 02215, USA}
\affiliation{Department of Astrophysics, St. Petersburg State University, Universitetsky pr. 28, Petrodvoretz, 198504 St. Petersburg, Russia}

\author[0000-0002-3638-0637]{Philip Kaaret}
\affiliation{University of Iowa Department of Physics and Astronomy, Van Allen Hall, 30 N. Dubuque Street, Iowa City, IA 52242, USA}
\affiliation{NASA Marshall Space Flight Center, Huntsville, AL 35812, USA}

\author[0000-0002-5760-0459]{Vladimir Karas}
\affiliation{Astronomical Institute of the Czech Academy of Sciences, Bo\v{c}n\'{i} II 1401/1, 14100 Praha 4, Czech Republic}

\author[0000-0001-7477-0380]{Fabian Kislat}
\affiliation{University of New Hampshire Department of Physics \& Astronomy Space Science Center Morse Hall, Rm 311 8 College Rd Durham, NH 03824}

\author{Takao Kitaguchi}
\affiliation{RIKEN Cluster for Pioneering Research, 2-1 Hirosawa, Wako, Saitama 351-0198, Japan}

\author[0000-0002-0110-6136]{Jeffery J. Kolodziejczak}
\affiliation{NASA Marshall Space Flight Center, Huntsville, AL 35812, USA}

\author[0000-0001-8916-4156]{Fabio La Monaca}
\affiliation{INAF Istituto di Astrofisica e Planetologia Spaziali, Via del Fosso del Cavaliere 100, I-00133 Roma, Italy}

\author[0000-0002-0984-1856]{Luca Latronico}
\affiliation{Istituto Nazionale di Fisica Nucleare, Sezione di Torino, Via Pietro Giuria 1, I-10125 Torino, Italy}

\author[0000-0001-9200-4006]{Ioannis Liodakis}
\affiliation{Finnish Centre for Astronomy with ESO, FI-20014 University of Turku, Finland}

\author[0000-0002-0698-4421]{Simone Maldera}
\affiliation{Istituto Nazionale di Fisica Nucleare, Sezione di Torino, Via Pietro Giuria 1, I-10125 Torino, Italy}

\author[0000-0002-0998-4953]{Alberto Manfreda}
\affiliation{Istituto Nazionale di Fisica Nucleare, Sezione di Pisa, Largo B. Pontecorvo 3, I-56127 Pisa, Italy}

\author[0000-0003-4952-0835]{Fr\'{e}d\'{e}ric Marin}
\affiliation{Universit\'{e} de Strasbourg, CNRS, Observatoire Astronomique de Strasbourg, UMR 7550, F-67000 Strasbourg, France}

\author[0000-0002-2055-4946]{Andrea Marinucci}
\affiliation{Agenzia Spaziale Italiana, Via del Politecnico snc, I-00133 Roma, Italy}

\author[0000-0001-7396-3332]{Alan P. Marscher}
\affiliation{Institute for Astrophysical Research, Boston University, 725 Commonwealth Avenue, Boston, MA 02215, USA}

\author[0000-0002-6492-1293]{Herman L. Marshall}
\affiliation{MIT Kavli Institute for Astrophysics and Space Research, Massachusetts Institute of Technology, 77 Massachusetts Avenue, Cambridge, MA 02139, USA}

\author[0000-0002-1704-9850]{Francesco Massaro}
\affiliation{Dipartimento di Fisica, Universit\`{a} degli Studi di Torino, Via Pietro Giuria 1, I-10125 Torino, Italy}
\affiliation{Istituto Nazionale di Fisica Nucleare, Sezione di Torino, Via Pietro Giuria 1, I-10125 Torino, Italy}

\author{Ikuyuki Mitsuishi}
\affiliation{Graduate School of Science, Division of Particle and Astrophysical Science, Nagoya University, Furo-cho, Chikusa-ku, Nagoya, Aichi 464-8602, Japan}

\author[0000-0001-7263-0296]{Tsunefumi Mizuno}
\affiliation{Hiroshima Astrophysical Science Center, Hiroshima University, 1-3-1 Kagamiyama, Higashi-Hiroshima, Hiroshima 739-8526, Japan}

\author[0000-0002-5847-2612]{C.-Y. Ng}
\affiliation{Department of Physics, The University of Hong Kong, Pokfulam, Hong Kong}

\author[0000-0002-1868-8056]{Stephen L. O'Dell}
\affiliation{NASA Marshall Space Flight Center, Huntsville, AL 35812, USA}

\author[0000-0002-5448-7577]{Nicola Omodei}
\affiliation{Department of Physics and Kavli Institute for Particle Astrophysics and Cosmology, Stanford University, Stanford, CA 94305, USA}

\author[0000-0001-6194-4601]{Chiara Oppedisano}
\affiliation{Istituto Nazionale di Fisica Nucleare, Sezione di Torino, Via Pietro Giuria 1, I-10125 Torino, Italy}

\author[0000-0001-6289-7413]{Alessandro Papitto}
\affiliation{INAF Osservatorio Astronomico di Roma, Via Frascati 33, I-00078 Monte Porzio Catone (RM), Italy}

\author[0000-0002-7481-5259]{George G. Pavlov}
\affiliation{Department of Astronomy and Astrophysics, Pennsylvania State University, University Park, PA 16802, USA}

\author[0000-0001-6292-1911]{Abel L. Peirson}
\affiliation{Department of Physics and Kavli Institute for Particle Astrophysics and Cosmology, Stanford University, Stanford, CA 94305, USA}

\author[0000-0003-3613-4409]{Matteo Perri}
\affiliation{Space Science Data Center, Agenzia Spaziale Italiana, Via del Politecnico snc, I-00133 Roma, Italy}
\affiliation{INAF Osservatorio Astronomico di Roma, Via Frascati 33, I-00078 Monte Porzio Catone (RM), Italy}

\author[0000-0003-1790-8018]{Melissa Pesce-Rollins}
\affiliation{Istituto Nazionale di Fisica Nucleare, Sezione di Pisa, Largo B. Pontecorvo 3, I-56127 Pisa, Italy}

\author[0000-0001-6061-3480]{Pierre-Olivier Petrucci}
\affiliation{Universit\'{e} Grenoble Alpes, CNRS, IPAG, F-38000 Grenoble, France}

\author[0000-0001-7397-8091]{Maura Pilia}
\affiliation{INAF Osservatorio Astronomico di Cagliari, Via della Scienza 5, I-09047 Selargius (CA), Italy}

\author[0000-0002-2734-7835]{Simonetta Puccetti}
\affiliation{Space Science Data Center, Agenzia Spaziale Italiana, Via del Politecnico snc, I-00133 Roma, Italy}

\author[0000-0003-1548-1524]{Brian D. Ramsey}
\affiliation{NASA Marshall Space Flight Center, Huntsville, AL 35812, USA}

\author[0000-0002-9774-0560]{John Rankin}
\affiliation{INAF Istituto di Astrofisica e Planetologia Spaziali, Via del Fosso del Cavaliere 100, I-00133 Roma, Italy}

\author[0000-0003-0411-4243]{Ajay Ratheesh}
\affiliation{INAF Istituto di Astrofisica e Planetologia Spaziali, Via del Fosso del Cavaliere 100, I-00133 Roma, Italy}

\author[0000-0002-7150-9061]{Oliver J. Roberts}
\affiliation{Universities Space Research Association (USRA)} 
\affiliation{NASA Marshall Space Flight Center, Huntsville, AL 35812, USA} 

\author[0000-0001-6711-3286]{Roger W. Romani}
\affiliation{Department of Physics and Kavli Institute for Particle Astrophysics and Cosmology, Stanford University, Stanford, CA 94305, USA}

\author[0000-0001-5676-6214]{Carmelo Sgr\'{o}}
\affiliation{Istituto Nazionale di Fisica Nucleare, Sezione di Pisa, Largo B. Pontecorvo 3, I-56127 Pisa, Italy}

\author[0000-0002-6986-6756]{Patrick Slane}
\affiliation{Center for Astrophysics, Harvard \& Smithsonian, 60 Garden Street, Cambridge, MA 02138, USA}

\author[0000-0002-7781-4104]{Paolo Soffitta}
\affiliation{INAF Istituto di Astrofisica e Planetologia Spaziali, Via del Fosso del Cavaliere 100, I-00133 Roma, Italy}

\author[0000-0003-0802-3453]{Gloria Spandre}
\affiliation{Istituto Nazionale di Fisica Nucleare, Sezione di Pisa, Largo B. Pontecorvo 3, I-56127 Pisa, Italy}

\author[0000-0002-2954-4461]{Douglas A. Swartz}
\affiliation{Universities Space Research Association (USRA)} 
\affiliation{NASA Marshall Space Flight Center, Huntsville, AL 35812, USA} 

\author[0000-0002-8801-6263]{Toru Tamagawa}
\affiliation{RIKEN Cluster for Pioneering Research, 2-1 Hirosawa, Wako, Saitama 351-0198, Japan}
\affiliation{
RIKEN Nishina Center, 2-1 Hirosawa, Wako, Saitama 351-0198, Japan}
\affiliation{Department of Physics, Tokyo University of Science, 1-3 Kagurazaka, Shinjuku, Tokyo 162-8601, Japan}

\author[0000-0003-0256-0995]{Fabrizio Tavecchio}
\affiliation{INAF Osservatorio Astronomico di Brera, Via E. Bianchi 46, I-23807 Merate (LC), Italy}

\author{Yuzuru Tawara}
\affiliation{Graduate School of Science, Division of Particle and Astrophysical Science, Nagoya University, Furo-cho, Chikusa-ku, Nagoya, Aichi 464-8602, Japan}

\author[0000-0002-9443-6774]{Allyn F. Tennant}
\affiliation{NASA Marshall Space Flight Center, Huntsville, AL 35812, USA}

\author[0000-0003-0411-4606]{Nicholas E. Thomas}
\affiliation{NASA Marshall Space Flight Center, Huntsville, AL 35812, USA}

\author[0000-0002-6562-8654]{Francesco Tombesi}
\affiliation{Dipartimento di Fisica, Universit\`{a}  degli Studi di Roma Tor Vergata, Via della Ricerca Scientifica 1, I-00133 Roma, Italy}
\affiliation{Istituto Nazionale di Fisica Nucleare, Sezione di Roma Tor Vergata, Via della Ricerca Scientifica 1, I-00133 Roma, Italy}
\affiliation{Department of Astronomy, University of Maryland, College Park, MD 20742, USA}

\author[0000-0002-3180-6002]{Alessio Trois}
\affiliation{INAF Osservatorio Astronomico di Cagliari, Via della Scienza 5, I-09047 Selargius (CA), Italy}

\author[0000-0002-9679-0793]{Sergey S. Tsygankov}
\affiliation{Department of Physics and Astronomy, FI-20014 University of Turku, Finland}

\author[0000-0002-4708-4219]{Jacco Vink}
\affiliation{Anton Pannekoek Institute for Astronomy \& GRAPPA, University of Amsterdam, Science Park 904, 1098 XH Amsterdam, The Netherlands}

\author[0000-0002-5270-4240]{Martin C. Weisskopf}
\affiliation{NASA Marshall Space Flight Center, Huntsville, AL 35812, USA}

\author[0000-0002-7568-8765]{Kinwah Wu}
\affiliation{Mullard Space Science Laboratory, University College London, Holmbury St Mary, Dorking, Surrey RH5 6NT, UK}

\author[0000-0002-0105-5826]{Fei Xie}
\affiliation{INAF Istituto di Astrofisica e Planetologia Spaziali, Via del Fosso del Cavaliere 100, I-00133 Roma, Italy}
\affiliation{Guangxi Key Laboratory for Relativistic Astrophysics, School of Physical Science and Technology, Guangxi University, Nanning 530004, People's Republic of
China}



\begin{abstract} 
Recent observations with the Imaging X-ray Polarimetry Explorer (\ixpe)  of two anomalous X-ray pulsars provided evidence that X-ray emission from magnetar sources is strongly polarized. Here we report on the joint \ixpe\ and \xmm\ observations of the soft $\gamma$-repeater \src. The spectral and timing properties of \src\ derived from \xmm\ data are in broad agreement with previous measurements; however, we found the source at an 
all-time-low persistent flux level. No significant polarization was measured apart from the $4$--$5\, \mathrm{keV}$ energy range, where  a probable detection with $\mathrm{PD}=31.6\pm10.5\%$ and $\mathrm{PA}=-17^\circ.6_{-15^\circ.0}^{+15^\circ.5}$ was obtained.
The resulting polarization signal, together with the upper limits we derive at lower and higher energies ($2$--$4$ and $5$--$8\, \mathrm{keV}$, respectively) is compatible with a picture in which thermal radiation from the condensed star surface is reprocessed by resonant Compton scattering in the magnetosphere, similar to what proposed for the bright magnetar \srcone. 

\end{abstract}

\keywords{X-rays: stars --- stars: magnetars ---
techniques: polarimetric}

\section{Introduction}
\label{sec:int}

Soft $\gamma$-repeaters (SGRs) and anomalous X-ray pulsars (AXPs) form together a small class of Galactic X-ray pulsars, characterized by long spin periods ($P\sim 1$--$12\, \mathrm s$), high spin-down rates $\dot P \sim 10^{-15}$--$10^{-10}\, \mathrm{s\,s^{-1}}$ and the emission of short, energetic bursts of hard X-/soft gamma-rays. The huge inferred values of the (spin-down) dipole field ($B\sim 10^{13}$--$10^{15}\, \mathrm G$), the lack of a detected binary companion and a persistent X-ray luminosity, $L_\mathrm{X} \sim 10^{30}$--$10^{35}\, \mathrm{erg\,s}^{-1}$, typically in excess of the spin-down power, indicate that these sources are magnetars, ultra-magnetized neutron stars powered by their own magnetic energy \citep[][see also \citealt{turolla+15,kaspi+belo17} for reviews and \citealt{ol+kaspi14} for a catalogue of magnetar sources\footnote{Available online at\\ \url{http://www.physics.mcgill.ca/~pulsar/magnetar/main.html}}]{td92,td93}. 

Because of the super-strong magnetic fields they host, the opacities of the two normal polarization modes are way different so the  X-ray emission from SGRs/AXPs is expected to be highly polarized \citep[up to $\approx 80\%;$][]{fernandez2011,taverna+14,taverna+20,Caiazzo2022}. Theoretical predictions were finally tested 
when the NASA-ASI  Imaging X-ray Polarimetry Explorer \cite[\ixpe;][]{Weisskopf2022}, the first satellite designed to provide imaging polarimetry in the $2$--$8\, \mathrm{keV}$ band, observed the two brightest magnetar sources, the AXPs \srcone\ and \srctwo\ (hereafter \srctwosh\ for short) during the first year of operations. Polarization was clearly detected in the $2$--$8$ keV band at the $\approx 13.5\%$ level in the former and to a much higher degree, $\approx 35\%$, in the latter \citep{taverna+22,zane+23}. 

In both sources the polarization strongly depends on energy. In \srctwosh\ it monotonically increases from $\approx 20\%$ up to $\approx 80\%$  at constant polarization angle, possibly indicating that $2$--$8\, \mathrm{keV}$ photons come from regions of the star surface with different properties: a magnetic condensate (either solid or liquid) and an atmosphere \citep[][]{zane+23}. On the other hand, in \srcone\ it first decreases from $\approx 15\%$ to zero at around $4$--$5\, \mathrm{keV}$  where the polarization angle swings by $90^\circ$, and then rises to $\approx 35\%$, suggesting that thermal radiation from a condensed surface patch is then reprocessed by resonant Compton scattering \citep[RCS;][]{tlk02,fernandez2007,ntz08} onto mildly relativistic electrons flowing in the star's twisted magnetosphere \citep[][an alternative interpretation in terms of mode switching in a magnetized atmosphere was recently put forward by \citealt{lai23}]{taverna+22}.

Polarization measurements in strongly magnetized neutron stars can probe vacuum birefringence, a strong-field, quantum electrodynamics (QED) effect predicted more than 80 years ago but never experimentally tested as yet \citep[see e.g.][and references therein]{heyl2000,heyl2002}. Although previous magnetar observations are in agreement with QED predictions, no smoking-gun evidence for vacuum birefringence was found, mostly because thermal emission in both \srcone\ and \srctwosh\ comes from a fairly limited area of the star surface, so its intrinsic polarization is preserved at infinity anyway, even if no vacuum birefringence is present \cite[see][]{vanAdelsberg2009,extp2019}. 

Further magnetar observations with \ixpe\ are required to address this issue and also to provide fresh insight into the similarities/differences among magnetar sources. Here we report on the simultaneous \ixpe\ and \xmm\ observations of the prototypical magnetar \src. The \xmm\ and \ixpe\ observations are detailed in \S\ref{sec:obs} and the results of the timing, spectral and polarimetric analyses are presented in \S\ref{sec:results}. Discussion follows in \S\ref{sec:discuss}.

\section{Observations} \label{sec:obs}

First identified as a high-energy transient in the KONUS data over 40 years ago \citep[][]{mazets+81}, \src\ was shortly after realized to be a repeater \citep[]{atteia+87,kouve+87,laros+87}. The source, located at about $8.7\,\mathrm{kpc}$ from the Sun \cite[][]{bibby+08}, is a regular and prolific emitter of short bursts clustered in active periods, one of which occurred in 2004 and culminated on 2004 December 27 with the emission of the most powerful giant flare (GF) observed so far from a magnetar 
\citep[$L\approx 10^{47}\, \mathrm{erg\,s}^{-1}$;][]{hurley+05,palmer+05}. 

\src\ spins with a period $P\approx 7.5\, \mathrm s$; the period derivative increased from  $\dot P= 8\times 10^{-11}\, \mathrm{s\,s^{-1}}$ to $5\times 10^{-10}\, \mathrm{s\,s^{-1}}$ in 2000--2011 and then went back to the ``historical'' value \cite[see][]{younes+17}. The latter implies a surface (dipole) field of $B= 8\times 10^{14}\, \mathrm{G}$, the highest ever recorded. The steady X-ray emission in the $0.5$--$10$ keV range, $L_\mathrm X\approx 10^{35}\, \mathrm{erg\,s^{-1}}$ (for the assumed distance of $8.7\, \mathrm{kpc}$), is well described by the superposition of a blackbody (BB) and a power law (PL) component, with (slightly variable) temperature $kT\approx 0.6\, \mathrm{keV}$, blackbody radius $\approx 1$--$2\, \mathrm{km}$ and spectral index $\Gamma\approx 1.6$ \citep[][]{mereghetti+05,woods+07,younes+15,younes+17}. The power law tail extends, seemingly unbroken, into higher energies  \citep[up to $\approx 200$ keV; see, e.g.,][]{mereghetti-int05,younes+17}. The pulse profile is double-peaked with a pulsed fraction $\mathrm{PF}\approx 3$--$8\%$ in a wide energy range \cite[][]{woods+07,younes+15,younes+17}.

The emission of a bright burst detected by several instruments on 2023, February 23 \citep[][]{GCN1,GCN2} marked the onset of a renewed period of activity from \src. To catch the source in an active state,  \ixpe\ observed \src\ starting on 2023, March 22. In addition a DDT pointing with \xmm\ was activated.

\subsection{\xmm}
\label{subsec:xmm}
\src\ was observed with the European Photon Imaging Camera (EPIC) on board the \xmm\ satellite starting on 2023-04-07 00:18:48 UTC for an exposure time of about $45\,\mathrm{ks}$. The EPIC-pn \citep{struder01} was operating in Full Frame mode (FF; timing resolution of $73.4\,\mathrm{ms}$). The MOS cameras \citep{turner01} were set in  Small Window mode (SW; timing resolution of $0.3\,\mathrm{s}$). 

Standard  procedures were applied in the extraction of the scientific products. Time intervals of high background activity were removed, resulting in a net exposure of $34.9\, \mathrm{ks}$ and $40.5\,\mathrm{ks}$ for the pn and the MOSs, respectively. We collected the source photons from within a circle of radius $42\arcsec$. The background level was estimated from a circular region of radius $100\arcsec$ centered far from the source, on the same CCD for the pn and, due to the reduced window, in a different CCD for the MOSs. We checked for the potential impact of pile-up with the \textsc{epatplot} tool and found a negligible pile-up fraction of $\approx 0.4\%$. The response matrices and ancillary files were generated by means of the {\sc rmfgen} and {\sc arfgen} tasks, respectively. The final source spectra were obtained by using {\sc specgroup}, rebinning the channels by a factor of three in order to match the intrinsic EPIC spectral resolution and imposing a minimum of $25$ counts in each channel.  Background-subtracted and exposure-corrected light curves were extracted using {\sc epiclccorr}. The EPIC photon arrival times were referenced to the Solar System barycenter. The source was found at a count rate of $0.354(3)\, \mathrm{cts\, s}^{-1}$ (here and in the following uncertainties are reported at $1\sigma$ confidence level, unless specified otherwise).

\subsection{\ixpe}
\label{subsec:ixpe}

\ixpe\ observed \src\ from 2023-03-22 05:57:51 UTC to 2023-04-01 18:56:26 UTC and from 2023-04-03 11:41:03 UTC to 2023-04-13 22:20:20 UTC, for a total on-source time of $\approx 947\, \mathrm{ks}$. 
Processed Level 2 (LV2) photon lists, one for each of the three \ixpe\ detector units (DUs), were downloaded from the archive at HEASARC\footnote{\url{https://heasarc.gsfc.nasa.gov/docs/ixpe/archive/}} and further processed to reduce the background with respect to the source signal. 

\begin{figure*}[tbh]
\includegraphics[width=\textwidth]{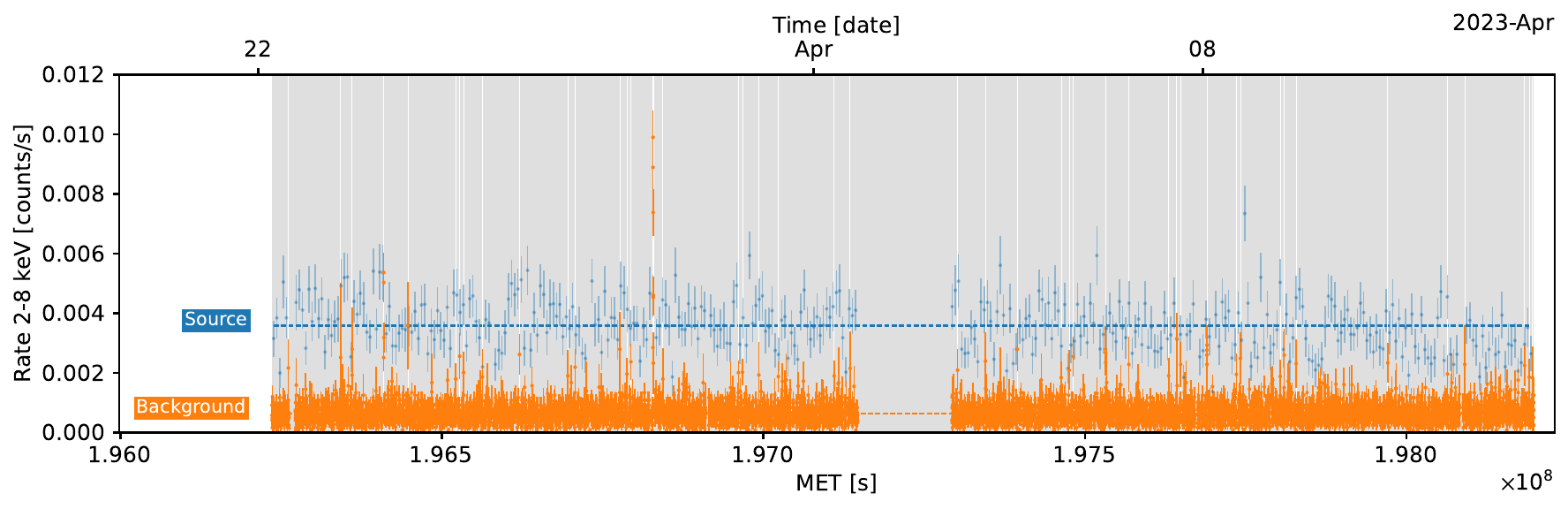} 
\caption{Count rate in the 2--8~keV energy range, summed over the three \ixpe\ DUs, as a function of time for two regions in the field of view, one containing the source and the other only background. Periods in which the background was larger than $50\%$ of the average source rate are identified by vertical white lines and removed by subsequent analyses. The time bin is $5000\, \mathrm{s}$ and $100\, \mathrm{s}$ for the source and background, respectively.
\label{fig:solar_bkg}}
\end{figure*}

Source and background counts were extracted from a circular region centered on the source position (identified with the brightest pixel in the \ixpe\ image) with radius $36\arcsec$ and from a concentric annulus with inner and outer radius of $78\arcsec$ and $240\arcsec$, respectively. In order to discriminate the background events due to charged particles and high-energy photons without affecting the genuine X-ray events, we then applied the rejection criteria described in \cite{DiMarco2023}.
The counting rate as a function of time is shown in Figure~\ref{fig:solar_bkg}. Time periods during which the background was larger than $50\%$ of the source average counting rate are flagged and removed from the subsequent analysis. This step removes $\approx 6$ out of the total $\approx 947\, \mathrm{ks}$ of the \ixpe\ observation. Arrival times were corrected to the Solar System barycenter with the FTOOL \textsc{barycorr} included in HEASOFT~6.31.1, using the Jet Propulsion Laboratory Development Ephemeris 421 and the International Celestial Reference System frame.

\section{Results} \label{sec:results}

\subsection{\xmm\ Timing and Spectral analysis}
\label{subsec:timing}

Both pn and MOS photon arrival time lists were used in order to look for the pulsar spin signal. We started from the timing properties inferred from the most recent spin measurements of \src\ in 2015--2016 \citep{younes+17} and extrapolated them to the epoch of the new \xmm\ dataset. In particular, we assumed a spin period $P=7.7501(2)\,\mathrm{s}$ and a first period derivative $\dot{P}=7.5(2)\times10^{-11}\,\mathrm{s\,s^{-1}}$, both referred to MJD 57202. The  linear evolution of the period in 2015--2016
(see Figure\,3 of \citealt{younes+17}) and the relatively quiet behaviour of the source in the last few years, suggest that $\dot{P}$ is constant (or at most slightly variable) since then. Correspondingly, we analyzed the new \xmm\ datasets searching for significant peaks in the $7.7659$--$7.7712\,\mathrm{s}$ period range, i.e. accounting for changes in $P$ and $\dot{P}$ up to about $6\sigma$ with respect to the values reported by \citet{younes+17}. 

\begin{figure*}[tbh]
\includegraphics[width=.45\textwidth]{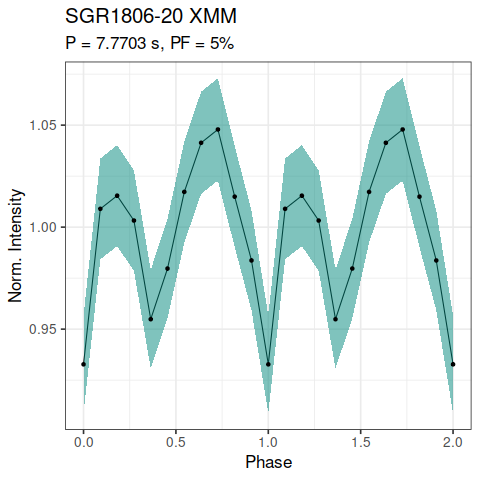} 
\includegraphics[width=.61\textwidth]{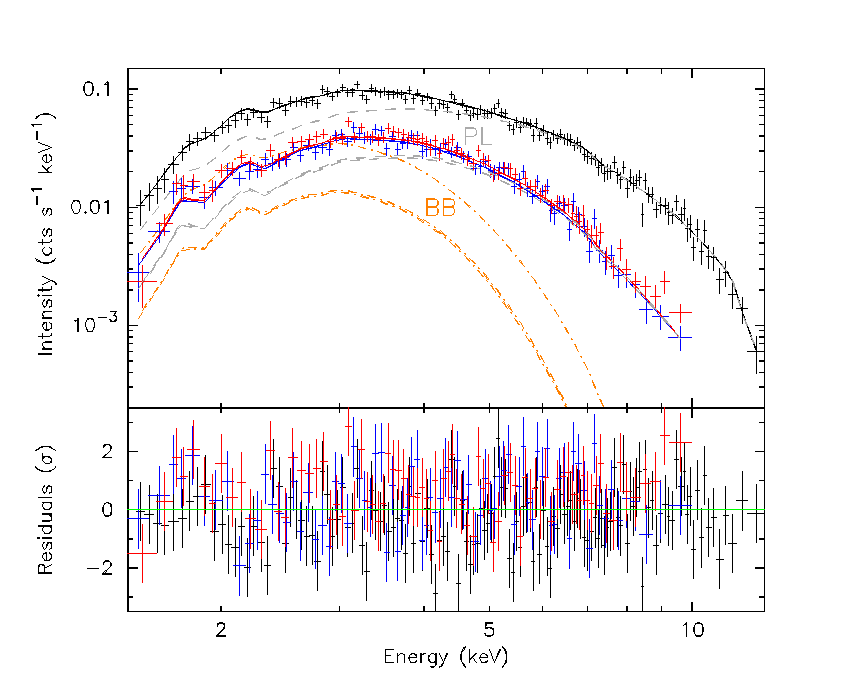} 
\caption{Left: The joint \xmm\ 1-12\,keV EPIC pn and MOS light curve folded to the best period inferred from the phase-fitting analysis (see text for details). The colored area marks the 1$\sigma$ uncertainties on the data points. Right: EPIC-pn (black crosses/lines), MOS1 (blue crosses/lines) and MOS2 (red crosses/lines) spectra of \src\ (upper panel), together with the fitting residuals, in units of standard deviation (lower panel), for the BB+PL model discussed in the text. The individual BB and PL components are shown by an orange dash-dotted and gray dashed line, respectively. 
\label{fig:xmmefold}}
\end{figure*}

Only one significant peak with a chance probability of about $9.6\sigma$ of not being a statistical fluctuation, and after having corrected for the number ($530$) of independent sampling periods of the search, was found in the Rayleigh periodogram. The period is $7.770\,\mathrm{s}$. The modulation is present both in the pn and in the MOS data alone, confirming that the signal is intrinsic to the source and not an instrumental artefact. In order to obtain a more refined measurement we applied a phase-fitting technique, which resulted in a period  of $7.7703(2)\,\mathrm{s}$ (frequency $\nu =0.128695(4)\,\mathrm{Hz}$); we derived also a $3\sigma$ upper limit on the period derivative, $|\dot{P}|<2\times 10^{-7}\, \mathrm{s\,s^{-1}}$. The $1$--$12\, \mathrm{keV}$ pulse shape is double-peaked, in agreement with the profiles obtained for \src\ in the past \citep{younes+15}. The pulsed fraction (defined as the semi-amplitude of the sinusoid divided by the source average count rate) is $\mathrm{PF} = 5(1)\%$ (see left panel of Figure\,\ref{fig:xmmefold}). 

\begin{table*}
\begin{center}
\caption{Results of the \xmm\ spectral fits.}
\label{tab:spec_ave}
\begin{tabular}{l|ccccccc}
\hline
Model & $N_{\mathrm H}$ & $kT_{{\rm BB}_1}$ & $R_{{\rm BB}_1}\,^a$ & $\Gamma\, \vert\, 
kT_{{\rm BB}_2}$ & Norm PL$^b\, \vert\, R_{{\rm BB}_2}\,^a$ & Flux$^c$ (Obs $\vert$ Unabs)  & $\chi^2$/dof. \\
   & ($10^{22}\, \mathrm{cm}^{-2}$)  &  (keV)  &   (km)  &  -- $\vert$ (keV)  &  -- $\vert$ (km)    & ($10^{-12}\, \mathrm{erg\,cm}^{-2}\mathrm s^{-1}$)& \\
\hline
BB+PL &6.5$\pm$0.3&0.59$\pm$0.04 & 1.3$^{+0.3}_{-0.2}$ & $1.7\pm 0.1$ & (1.3$\pm$0.3)$\times$10$^{-3}$ & 4.3$\pm 0.1\,\vert\, 11.3\pm 0.4$ & 351.9/298 \\ 

BB+BB & 5.7$\pm$0.2 &0.70$\pm$0.03 & 1.2$\pm$0.1 & 2.4$\pm$0.1 & 0.11$\pm$0.01 & 4.3$\pm 0.1\,\vert\,  8.2 \pm 0.2$  & 351.2/298 \\

\hline
\hline
\end{tabular}
\begin{list}{}{}
\item[$^a$] Derived by adopting a 8.7 kpc distance \citep{bibby+08}.
\item[$^b$] In units of $\mathrm{counts\,keV}^{-1}\,\mathrm{cm}^{-2}\,\mathrm{s}^{-1}$.
\item[$^c$] The fluxes are estimated in the $0.5$--$10\,\mathrm{keV}$ energy range.
\end{list}
\end{center}
\end{table*}

The spectral analysis of \src\ was performed by fitting simultaneously all the EPIC datasets within XSPEC \cite[][]{Arnaud1996}. A single component (absorbed) model, such as a PL or a BB, resulted in a poor fit (reduced $\chi^2\approx1.3$--$3.1$ for 300 dof). A BB+PL and a BB+BB model, often adopted to
describe the $0.1$--$12\, \mathrm{keV}$ magnetar spectra, yield a good agreement with the data (reduced $\chi^2\approx$ 1.17 in both cases). The best-fitting parameters are listed in Table\,\ref{tab:spec_ave} and the EPIC spectra together with the best-fitting BB+PL model are shown in the right panel of Figure\,\ref{fig:xmmefold}. We find that the addition of a second spectral component is significant to more than $6\sigma$. Although the BB+BB and BB+PL models are equally acceptable on a statistical basis, in the following we refer to the latter. The BB+PL spectral decomposition has been widely adopted in the past for \src, and this choice  is motivated by the clear detection of a power law tail e.g. in \nustar\ data \citep{younes+15,younes+17}. The inferred parameters are in agreement with the average values obtained in the past ten years, though the flux was unexpectedly a factor of about three lower than the average. 

\subsection{\ixpe\ timing and spectro-polarimetric analysis}
\label{subsec:polspec}

After background and solar-flare removal (see \S\ref{subsec:ixpe}), no significant changes in both the source and background counts were detected in the two \ixpe\ pointings, so that we used the combined data in our analysis.

We searched for pulsations using the $Z^2_2$ statistics \citep{Buccheri1983}, which is adequate for double-peaked pulsed profiles.
We joined the event lists from the three DUs and ran the search using the quasi-fast folding algorithm described in \citet{bachettiAllOnceTransient2020,bachettiExtendingStatisticsGeneric2021}. 
We searched over spin frequencies between $0.125$ and $0.129\, \mathrm{Hz}$, that contained the \xmm\ solution, and frequency derivatives $|\dot{\nu}_{\rm spin}|< 4 \times 10 ^{-12}\,\mathrm{Hz\,s^{-1}}$. Given the low number of source counts (see below),
we did not find significant peaks in the search; we derived a $5.2\%$ pulsed fraction upper limit on pulsations over the frequency interval indicated above ($Z^2_2\approx25$; $90\%$ c.l., evaluated following \citealt{vaughanSearchesMillisecondPulsations1994}).  This upper limit is compatible with the detection with \xmm\ of a $\approx 5\%$ pulsed fraction.

We extracted the Stokes parameters $I$, $Q$ and $U$ with the {\tt xpbin} tool, exploiting the weighted analysis method \cite[][]{DiMarco22} implemented in the latest version (30.3) of the \ixpeobssim\ software \cite[][]{Baldini2022}\footnote{\url{https://github.com/lucabaldini/ixpeobssim}}. As a consequence of the limited number of counts ($\approx 8000$ background-subtracted events in the three DUs), the phase- and energy-integrated ($2$--$8$ keV) polarization degree, $\mathrm{PD}$$=$$\sqrt{Q^2+U^2}/I$, is $5.7\%$, below the $\mdp$ \cite[][]{Weisskopf2010}, which is about $20\%$ for this observation and is therefore not significant\footnote{The minimum detectable polarization ($\mathrm{MDP}$) is the largest signal expected to be produced by statistical fluctuations only, at a given confidence level (usually $99\%$, $\mdp$). The result of a polarization measurement is regarded as significant if it exceeds $\mdp$.}. The polarization angle, $\mathrm{PA}$$=\arctan{(U/Q)}/2$, is then unconstrained. The same conclusion is reached by applying an analysis as outlined by \citet{Strohmayer2017} within XSPEC, which provides $\mathrm{PD}\approx 4.7\%$.

In order to test if a detectable polarization is present in some specific energy intervals, we performed an energy-resolved, phase-integrated (weighted) analysis in XSPEC. We started by dividing the \ixpe\ working energy band into $6$ equal bins (each $1$-keV wide). The null hypothesis probability that the source is unpolarized  in all the energy bins is $\approx 22\%$. 

We found a signal with $\mathrm{PD}=31.6\pm10.5\%$ (slightly higher than the $\mdp$) and $\mathrm{PA}=-17^\circ.6_{-15^\circ.0}^{+15^\circ.5}$, computed East of North, in the $4$--$5\, \mathrm{keV}$ range. No significant polarization was detected in the remaining bins.
In particular, we found only an upper limit of $23\%$ and $45\%$ ($99\%$ confidence level) in the two neighbouring bins, $3$--$4$ and $5$--$6\,\mathrm{keV}$, respectively. Despite no firm conclusion can be reached about the trend of PD with energy, this may suggest that it increases in going  from $3$ to $4$--$5\,\mathrm{keV}$. What happens at higher energies is harder to tell, essentially because the lower S/N ratio makes the $\mdp$ higher and this translates into a weaker constraint on PD.   

\begin{figure*}[!tbh]
\includegraphics[trim={0 2cm 0 2cm},clip]{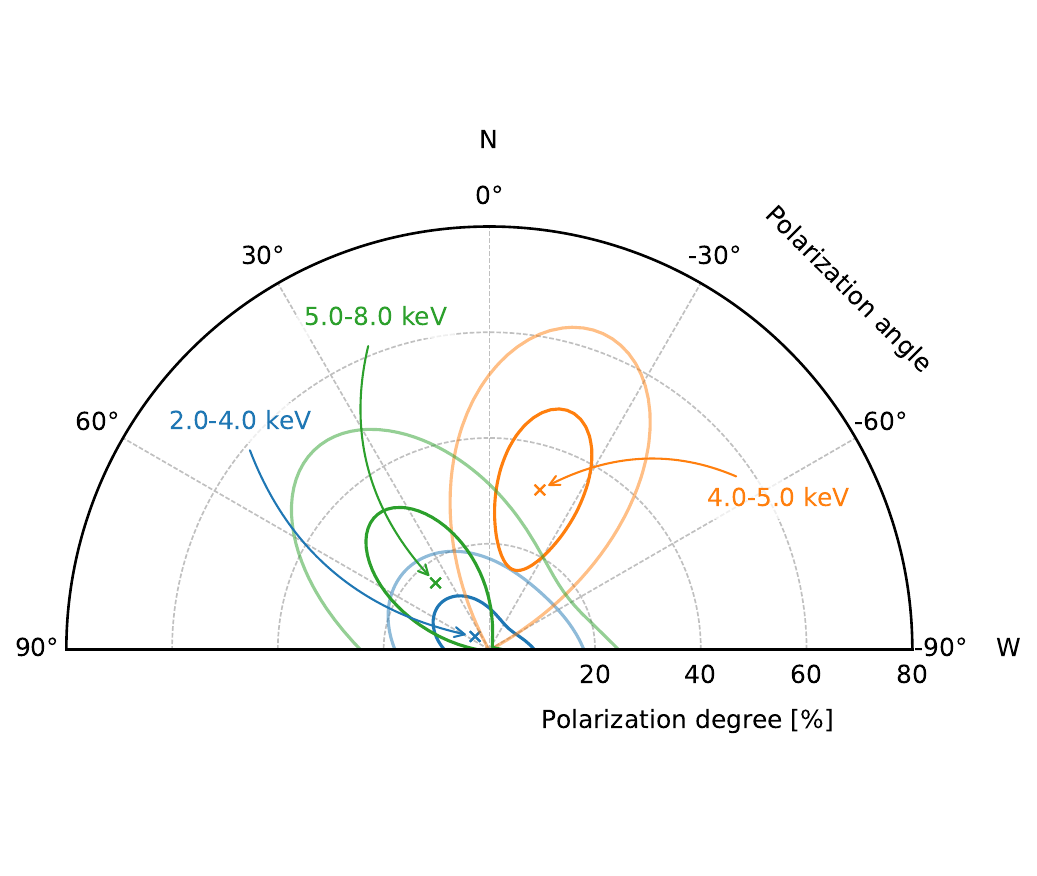} 
\caption{Contour plot of the \ixpe\ polarization degree and angle in different energy bands. The inner and outer contours mark the regions at $68.3\%$ and $99\%$ confidence level, respectively. The spectral model is frozen to that obtained from the \xmm\ observation. \label{fig:contours}}
\end{figure*}

In order to improve the counting statistics we merged together the first and last pair of bins and considered the intervals $2$--$4$, $4$--$5$ and $5$--$8\,\mathrm{keV}$. This did not produce any improvement, resulting again in an upper limit for $\mathrm{PD}$ of $24\%$ and of $55\%$ ($99\%$ confidence level) at low and high energies, respectively. 
Figure~\ref{fig:contours}
shows the contour plots of the polarization degree and angle in each energy band, obtained with the XSPEC {\tt steppar} command by assuming the spectral model obtained from the \xmm\ observation. A somewhat different choice for the boundaries of the central interval, e.g. taking the bin from $3.5$ to $5\,\mathrm{keV}$ or from $4$ to $5.5\,\mathrm{keV}$, yields consistent results. 


We also attempted a phase-dependent analysis by folding the data at the spin period derived from the \xmm\ timing and assuming $\dot{P}=7.5\times10^{-11}\,\mathrm{s\,s^{-1}}$ (see \S\ref{subsec:timing}). No significant polarization was detected after dividing the pulse cycle the into 7 equally-spaced phase bins. In the next step we restricted to two phase bins, corresponding to the intervals $[0.0,\,0.38]$ and $[0.38,\,1.0]$, selected in such a way as to contain the secondary and primary peak of the pulse, respectively. Again no polarization was measured, either considering the entire $2$--$8\, \mathrm{keV}$ band or restricting to low ($2$--$4$ keV) and high ($4$--$8$ keV) energies. A negative result was also found limiting the analysis to the  $4$--$5\, \mathrm{keV}$ range, where a phase-integrated signal was detected at $99\%$ confidence level. This was expected given that the detection is marginal even when all the counts are considered.

An unbinned analysis \cite[without background subtraction;][]{gonzalez23} in the $2$--$8$ keV range shows that, when marginalized over the various parameters, the median polarization degree is  $\approx 8.4\%$ and about $1\sigma$ above zero. This is somewhat less than the $\mathrm{MDP}_{99}$ of $10\%$ for the observation without binning in phase or energy.

Fits to \ixpe\ count spectra are rather inconclusive. Single component models, either an (absorbed) BB or PL, provide a (formally) acceptable agreement with the data. However, the BB fit yields a column density of $2.7\times 10^{22}\, \mathrm{cm}^{-2}$, more than twice lower than that measured by \xmm\ and previously reported in the literature. On the other hand,
the parameters derived from the PL fit are consistent (within $\approx 1\sigma$) with those from the \xmm\ BB+PL fit, but this just reflects the limited \ixpe\ energy range ($2$--$8\, \mathrm{keV}$). When a single PL model is extended to the entire \xmm\ energy band the fit is no longer satisfactory (see \S \ref{subsec:timing}). A fit with a two component model (BB+BB or BB+PL) leaving all the parameters free to vary is largely unconstrained, although by freezing all the parameters (except the normalizations) to those derived from the \xmm\ data provides a good fit ($\chi^2=129.8$ for $162$ dof). 

A joint fit of the three Stokes parameters with the XSPEC model {\tt phabs$\times$
(bbodyrad$\times$polconst} {\tt $+$powerlaw$\times$polconst}), freezing again all the spectral parameters to those of the \xmm\ analysis, does not provide any conclusive result. The upper limit ($3\sigma$ confidence level) on the polarization degree of the PL is about $57\%$ while the polarization of the BB is unconstrained. A similar conclusion is reached by truncating the PL at low energies. We remark that, while the fit is statistically acceptable in all these cases, nothing can be said about the relative polarization direction between the two components.


\section{Discussion }\label{sec:discuss}

\ixpe\ observed \src\ in two segments for a total exposure time of about $1\, \mathrm{Ms}$ in March--April 2023. During the second stint, \xmm\ targeted the source for about $45\, \mathrm{ks}$ to provide complementary spectral and timing information. Despite the renewed bursting activity, \src\ was found at a (unabsorbed) flux level of $ 1.1\times 10^{-11}\, \mathrm{erg\,cm^{-2}\,s^{-1}}$ in the $0.5$--$10\,\mathrm{keV}$ band, the lowest ever recorded, although still compatible within the uncertainties with the \nustar\ one reported by \cite{younes+17}. \xmm\ spectra are well fitted by an (absorbed) BB+PL model and the  spectral parameters ($kT_\mathrm{BB}= 0.6\, \mathrm{keV}$, $R_\mathrm{BB}= 1.3\, \mathrm{km}$, $\Gamma= 1.7$) are in broad agreement with those found by \cite{younes+17}. There is, however, a hint for a decrease of the BB radius with respect to the values measured by \cite{younes+17} in 2015--2016, even if their \nustar\ data did not allow for a precise determination of the BB flux, and of a steepening of the PL with respect to the values measured by \xmm\ in the same energy range prior to 2011 \citep[][]{younes+15}.  The period and the upper limit for $\dot P$ we found are consistent with the past timing history of the source. The \xmm\ pulse profile is double peaked with a pulsed fraction $PF\approx 5\%$, in agreement with previous measurements, while no pulsations were detected in the \ixpe\ data.

The low flux level of \src\ and the high background prevented a complete spectro-polarimentric and timing analysis of \ixpe\ data. No pulsations were detected, with an upper limit of $5.2\%$ on the pulsed fraction. Although poorly constrained, the \ixpe\ spectrum is compatible with the BB+PL decomposition obtained from the analysis of \xmm\ data. No significant polarization has been detected integrating over the source rotational period and in the $2$--$8$ keV energy band. 

Interestingly, by restricting the analysis to the $4$--$5$ keV range we found a polarization signal significant at $99\%$ confidence level, with PD $\approx32\%$ and PA $\approx -18^\circ$. At lower ($2$--$4$ keV) and higher ($5$--$8$ keV) energies only upper limits can be derived on PD, with values at $3\sigma$ of $24\%$ and $55\%$, respectively.

Although \xmm\ spectra are compatible with both a BB+PL and a BB+BB decomposition, the former is in our opinion favored, as discussed in \S\ref{subsec:timing}. If the power law tail is associated with RCS,
the predicted polarization degree saturates at $33\%$ in the energy interval where this component dominates \cite[][]{taverna+20}. The thermal emission can originate either from a magnetized, cooling atmosphere, with PD $\approx70$--$80\%$ \cite[except for very peculiar viewing geometries, see e.g.][]{taverna+15}, or from the bare, condensed surface of the star, in which case a much lower PD $\lesssim15$--$20\%$ is expected. It has been also proposed that radiation emerging from atmospheres heated from above by particle bombardment has a modest polarization degree \citep[comparable to that of the condensed surface,][]{gonzalez2019,dorosh+22}.

The constraints placed by the upper limits at low and high energies, together  with a $99\%$-c.l. polarization degree of $\approx 32\%$ at intermediate energies (where the thermal and non-thermal component coexist, see above) are compatible with a picture in which thermal emission comes from the bare NS surface or from an atmosphere heated by backflowing particles. This scenario was also applied to \srcone, where a clear $90^\circ$ swing of PA with energy was detected, signalling that radiation is dominated by O-mode photons (coming from the condensed surface/heated atmosphere) at low energies and by X-mode photons (reprocessed by scatterings) at higher ones. The two sources may therefore be similar. Due to the low signal-to-noise ratio, no conclusion can be drawn about the dependence of the polarization direction on the energy and hence on which polarization mode is prevailing in a given energy range. However, emission from a condensed surface/heated atmosphere can be either mostly in the X- or in the O-mode \cite[depending e.g. on the orientation of the local magnetic field and the photon energy for the former and on the temperature gradient for the latter, see][]{taverna+22,zane+23}, so that a swing in PA is not necessarily expected.

Along this line, we explored a simple RCS model, in which thermal emission comes from two hot spots placed near the magnetic equator of the bare NS surface, including vacuum birefringence. With this choice of the emission geometry (similar to that adopted for \srcone), radiation from the magnetic condensate exhibits a relatively large polarization degree ($\approx 20\%$ in the O-mode) at low energies \cite[$\lesssim 4\, \mathrm{keV}$;][]{taverna+22}. Since the magnetic field in the outer layers of a magnetar is expected to locally deviate from a dipole, the surface thermal map is likely different from the usual (dipolar) one, in which the hotter regions are around the magnetic poles, and may change in time \cite[see][for some observational evidences; see also \citealt{degrandis+20, degrandis+21} for theoretical considerations]{tiengo+13,borghese+22}.  Such a configuration indeed reproduces quantitatively both the observed \xmm\ $1$--$10$ keV spectrum and the $2$--$8$ keV pulse profile, as well as being
compatible with the \ixpe\ detection/upper limits of PD in different energy bins (see Figure \ref{fig:contours}).  

The low pulsed fraction of \src\ points to a sizable emitting surface area, as further supported by the inferred BB radius, $\approx 2\,\mathrm{km}$, rather large compared to that usually seen in other magnetar sources \cite[$R_\mathrm{BB}\approx 0.1$--$1\, \mathrm{km}$;][see references therein]{ol+kaspi14}. The actual size of the emitting region depends on the emission properties of the surface (and also on the geometry and viewing angle of the source). For poorly radiating surfaces, like a magnetic condensate, it is typically larger than $R_\mathrm{BB}$. In the model outlined above, the (total) size is $\approx 3.5\, \mathrm{km}$, for an assumed star radius $R_\mathrm{NS}=13\, \mathrm{km}$. The size of the emitting region is not per se a primary factor in establishing the intrinsic polarization properties of the source, which are mostly determined by the physical processes occurring in the surface/magnetosphere and by the source geometry. Indeed, by replacing in our model the magnetic condensate with a fully ionized, magnetized, H atmosphere produces too a large polarization degree at lower energies. On the other hand, the extent to which QED affects the observed polarization depends on the size of the emitting area. If radiation comes from a limited region, across which the direction of the $B$-field changes little, vacuum birefringence produces almost no detectable effect, i.e. the predicted polarization at infinity is the same with or without QED. In case the emitting area has size $\approx R_\mathrm{NS}$, instead, the polarization computed without QED effects is quite lower than that with QED. Despite \src\ appears promising in this respect, the low counting statistics prevented us to test the QED vs. no-QED scenario. 

The quite large upper limit above $\approx 5\,\mathrm{keV}$ does not rule out different emission scenarios in which the PD is actually larger than $\sim 30\%$, the value predicted by saturated RCS. 
Phase-averaged values as high as $\sim 55\%$, for instance, may hint at the presence of hot gaseous caps on the star surface, as proposed by \cite{zane+23} for \srctwosh. This possibility would  be still consistent with the spectral properties of the source, given that a BB+BB model also provides a good representation of \xmm\ spectra.  

 \ixpe\ observations of magnetars have clearly demonstrated the capabilities of polarization measurements in probing the physical conditions in strongly magnetized neutron stars. This includes the presence of QED effects, in favor of which indirect evidences were already gathered from the analysis of \srcone\ and \srctwosh. The emerging picture shows substantial differences in the polarization pattern between \srcone\ (and possibly \src) and \srctwosh, signalling that emission at higher energies ($\sim 4$--$8\, \mathrm{keV}$) may arise from different physical conditions in the different sources. As the case of \src\ shows, collecting enough counts is crucial in providing clear-cut results. Further progress will require observing a larger sample of weaker sources which will be possible for \ixpe\ only with longer exposure times.

\begin{acknowledgments}
The Imaging X-ray Polarimetry Explorer (\ixpe) is a joint US and Italian mission.  The US contribution is supported by the National Aeronautics and Space Administration (NASA) and led and managed by its Marshall Space Flight Center (MSFC), with industry partner Ball Aerospace (contract NNM15AA18C).  The Italian contribution is supported by the Italian Space Agency (Agenzia Spaziale Italiana, ASI) through contract ASI-OHBI-2017-12-I.0, agreements ASI-INAF-2017-12-H0 and ASI-INFN-2017.13-H0, and its Space Science Data Center (SSDC) with agreements ASI-INAF-2022-14-HH.0 and ASI-INFN 2021-43-HH.0, and by the Istituto Nazionale di Astrofisica (INAF) and the Istituto Nazionale di Fisica Nucleare (INFN).  This research used data products provided by the \ixpe\ Team (MSFC, SSDC, INAF, and INFN).  
R. Tu., R. Ta. and G.L.I. acknowledge financial support from the Italian MUR through grant PRIN 2017LJ39LM. G.L.I. also acknowledges financial support from INAF, through grant “IAF-Astronomy Fellowships in Italy 2022 - (GOG)”. 
J.H. acknowledges support from the Natural Sciences and Engineer Council of Canada and the Canadian Space Agency.
M.Negro acknowledges the support by NASA under award number 80GSFC21M0002.T.T. was supported by grant JSPS KAKENHI JP19H05609. HK and EG acknowledge NASA support under grants 80NSSC18K0264, 80NSSC22K1291, 80NSSC21K1817, and NNX16AC42G. 
We thank N. Schartel  for granting us \xmm\  Director Discretionary Time.

\end{acknowledgments}

\facilities{\ixpe, \xmm}





%
\bibliography{1806}{}
\bibliographystyle{aasjournal}
\end{document}